\documentclass[journal]{IEEEtran}

\usepackage{xcolor,soul,framed} 
\colorlet{shadecolor}{yellow}
\usepackage[pdftex]{graphicx}
\graphicspath{{../pdf/}{../jpeg/}}
\DeclareGraphicsExtensions{.pdf,.jpeg,.png}

\usepackage{cite}
\usepackage[cmex10]{amsmath}
\usepackage{amssymb}
\usepackage{array}
\usepackage{mdwmath}
\usepackage{mdwtab}
\usepackage{multirow}
\usepackage{bbm}
\usepackage{breqn}
\usepackage{float}
\usepackage{eqparbox}
\usepackage{url}
\usepackage{caption}
\usepackage{tabularx,makecell,hhline}
\usepackage[caption=false]{subfig}
\usepackage{algorithmic,algorithm,xpatch}
\begin{document}
\bstctlcite{IEEEexample:BSTcontrol}
    \title{kPF-AE-LSTM: A Deep Probabilistic Model for  Net-Load Forecasting in High Solar Scenarios}
  \author{Deepthi~Sen, Indrasis Chakraborty, Soumya Kundu, Andrew P. Reiman, Ian Beil, and Andy Eiden\\
  \thanks{D.\,Sen is with the Quantum Energy Partners, TX, USA (and was with Lawrence Livermore National Laboratory, CA, USA, during this work), email: sen.deepthi@gmail.com. I.\,Chakraborty is with the Lawrence Livermore National Laboratory, CA, USA, email: chakraborty3@llnl.gov. S.\,Kundu holds a joint appointment with the Pacific Northwest National Laboratory, WA, USA, and University of Vermont, VT, USA, email: soumya.kundu@pnnl.gov. A.\,P.\,Reiman is with the Pacific Northwest National Laboratory, WA, USA, email: andrew.reiman@pnnl.gov. I.\,Beil and A.\,Eiden are with the Portland General Electric, OR, USA, email: \{ian.beil,\,andy.eiden\}@pgn.com.}}


\maketitle

\begin{abstract}
With the expected rise in behind-the-meter solar penetration within the distribution networks, there is a need to develop  time-series forecasting methods that can reliably predict the net-load, accurately quantifying its uncertainty and variability. This paper presents a deep learning method to generate probabilistic forecasts of day-ahead net-load at 15-min resolution, at various solar penetration levels. Our proposed deep-learning based architecture utilizes the dimensional reduction, from a higher-dimensional input to a lower-dimensional latent space, via a convolutional Autoencoder (AE). The extracted features from AE are then utilized to generate probability distributions across the latent space, by passing the features through a kernel-embedded Perron-Frobenius (kPF) operator. Finally, long short-term memory (LSTM) layers are used to synthesize time-series probability distributions of the forecasted net-load, from the latent space distributions. The models are shown to deliver superior forecast performance (as per several metrics), as well as maintain superior training efficiency, in comparison to existing benchmark models. Detailed analysis is carried out to evaluate the model performance across various solar penetration levels (up to 50\%), prediction horizons (e.g., 15\,min and 24\,hr ahead), and aggregation level of houses, as well as its robustness against missing measurements. 
\end{abstract}


%
\IEEEpeerreviewmaketitle


\section{Introduction}

\IEEEPARstart{R}{eliable} and accurate means of forecasting power generation and net-load are essential for maintaining an uninterrupted energy supply in the power grid. Forecasting becomes especially valuable in the context of utilization of renewable power, which is inherently dependent on atmospheric conditions and therefore uncertain. The uncertainty associated with renewable power sources such as wind and solar also increases the energy cost by requiring ancillary frameworks for load balancing during periods of unexpectedly high and low production~\cite{lueken2012costs}. The need for a high-accuracy forecasting framework for renewable power was identified in as early as the 1980s and early 90's in the context of wind~\cite{bodin1980uncertainty}, general load forecasting~\cite{yunwen1985load,hongbo1994application} and solar power~\cite{badrul}. Since then, a large body of literature has explored a wide range of methods, from theoretical modelling to statistical approaches, for forecasting electrical load and renewable power generation. 

A recent article \cite{alkhayat2021review} provides a comprehensive review of various forecasting approaches developed for renewable power sources such as solar and wind. Some of the advanced solar forecast tools involve satellite imagery, numerical weather prediction (NWP) and cloud tracking \cite{jang2016solar,blanc2017short,cros2014extracting,chow2011intra}. This is inspired by the fact that solar production is known to be a strong function of the cloud-related parameters including cloud index, type, and height, as well as relative motion~\cite{kleissl2013solar}. 
However, meteorological information at such level of detail is rarely available, and  this motivates the development of forecasting models that rely on more readily available data such as air temperature and relative humidity. The works in \cite{GOH,sharif} used conventional time-series analysis models (e.g., autoregressive moving average, 
autoregressive integrated moving average) 
to forecast solar irradiation data.
The work in \cite{chu2015short} compared three forecasting models, based on cloud-tracking, time-series analysis, 
and $k$-nearest neighbours. An artificial neural network model was used to re-predict the outcomes of the prior model to achieve better prediction for 5-10 mins ahead. The authors of \cite{abedinia} proposed an amalgamation of a neural network, a metaheuristic algorithm, and a feature selection procedure that discards irrelevant data and improves the neural network training. In this work, they used solar irradiance, temperature, and historical data to generate the forecasts, and tested the algorithm on a 8-day test period.
%
The article \cite{golestaneh} proposed an extreme-machine-learning (ELM) based approach to obtain a probabilistic forecast of solar power, in the form of both point and percentile estimates, from a wide array of inputs such as clear sky solar irradiance,  air temperature, wind speed, humidity, and solar irradiance. 

\subsection{Recurrent Neural Networks in Solar Forecasting}
However, the work in \cite{hossain} showed that a recurrent neural network (RNN) based forecasting model offered better performance than the ELM in terms of mean absolute percentage error (MAPE). Furthermore, the work in \cite{alkhayat2021review} reported that RNNs are among the most utilized models for solar and wind forecasting. The authors of \cite{zhou2019short} used long-short-term-memory (LSTM) networks -- a type of RNN -- with attention mechanism for short term (up to 60 min) solar power forecasts and achieved $\sim$20\% MAPE. 
A notable improvement that LSTMs introduce over simple RNN is the ease of optimization, which is made possible with the use of gates, for storing and retrieving long-term information. However, the model in \cite{zhou2019short} is deterministic and the authors have not reported the spread of errors to quantify the reliability and generalizability of the model. A number of other works, such as \cite{qing2018hourly,zhang2018solar,yu2019lstm}, have explored the application of various LSTM-based architectures. 
In a recent work \cite{wang2018wavelet}, the authors proposed a convolutional layer to extract important features from the input window prior to an LSTM cell, thereby improving the training efficiency, and applied the concept to solar irradiation prediction. However, the predictions were deterministic and the uncertainty in the estimates was not studied. In \cite{doubleday2020benchmark}, the authors compared five different benchmark probabilistic methods for an hourly solar forecasting task. But they did not consider the problem of net-load forecasting in presence of behind-the-meter solar.

\subsection{Existing Literature on Net-Load Forecasting}

Compared to solar forecasting problem, there are relatively fewer studies on net-load forecasting under high behind-the-meter (BTM) solar penetration \cite{ALIPOUR2020118106}. Broadly speaking, there are two ways to forecast the net-load under high BTM solar penetration. The first involves directly forecasting net-load and the second requires separate forecast models for solar generation and demand, which are aggregated to form the net-load forecast. It was argued in \cite{kaur2016net,van2018review}, that an integrated approach of directly synthesizing the net-load forecast via systematic combination of both solar and end-use demand variability can significantly enhance the forecast accuracy. The work in \cite{ALIPOUR2020118106} used wavelet transforms to decompose the net-load time-series into several time-series of low and high frequencies. Subsequently, auto-encoders were used to extract effective features for prediction. In \cite{sun}, the authors used Bayesian deep learning to generate probabilistic residential net-load forecasts using LSTMs. They used solar PV production measurements from houses with solar installations to improve the net-load forecast. The models were trained by minimizing the uncertainty in the prediction, accounting for both the inherent and model uncertainty. Authors of \cite{wang} proposed a probabilistic net-load forecasting framework wherein the net-load was decomposed into the actual load, solar PV output and the residual. Subsequently, equivalent PV parameters were computed and a correlation analysis carried out. 
Another approach introduced by \cite{kobylinski2020high} enables high resolution net load forecasting for neighbourhoods with high penetration of renewable power, using an artificial neural network-based model with a special emphasis on feature selection. 
In addition to the aforementioned works, \cite{CORIZZO,corizzo2019dencast,buhan2015multistage,cavalcante2017lasso,ko2020deep} explore a wide variety of approaches for reliably forecasting net-load and various kinds of renewable energy not limited to solar power. These approaches range from physics-based modeling and NWP to completely data-driven ones such as naive Bayesian methods, LASSO vector auto-regression and density-based clustering.

\subsection{Main Contributions and Overall Structure}
We propose a novel deep probabilistic architecture for time-series forecasting of net-load under uncertainties caused by high behind-the-meter solar penetration. The following are the main contributions of this work:
\begin{enumerate}
    \item We propose a novel deep learning model, henceforth referred to as kPF-AE-LSTM, consisting of two key components: a kernel-embedded Perron-Frobenius operated autoencoder and a probabilistic LSTM, to reliably capture the uncertainties and variability in net-load forecast under high solar penetration.
    \item We benchmark the proposed model against state-of-the-art deep probabilistic models, to demonstrate best-in-class forecast accuracy and training efficiency.
    \item We perform extensive analysis to demonstrate the reliability and robustness of the model performance under varying solar penetration (up to 50\%), varying level of aggregation, and missing measurements.
\end{enumerate}
The rest of this paper is structured as follows. In Sec.\,\ref{sec2}, we describe the dataset used to evaluate our proposed architecture. We describe the unique features of the proposed forecasting architecture, and the associated training process, in Sec.\,\ref{sec3}. Detailed performance evaluation and benchmarking of the architecture under varying scenarios (including, solar penetration, aggregation of houses, and missing measurements) are presented in Sec.\,\ref{sec4}. We conclude the paper in Sec.\,\ref{sec5}.

\section{Dataset Description}
\label{sec2}

\begin{figure}[!ht]
    \centering
    \includegraphics[width=\columnwidth]{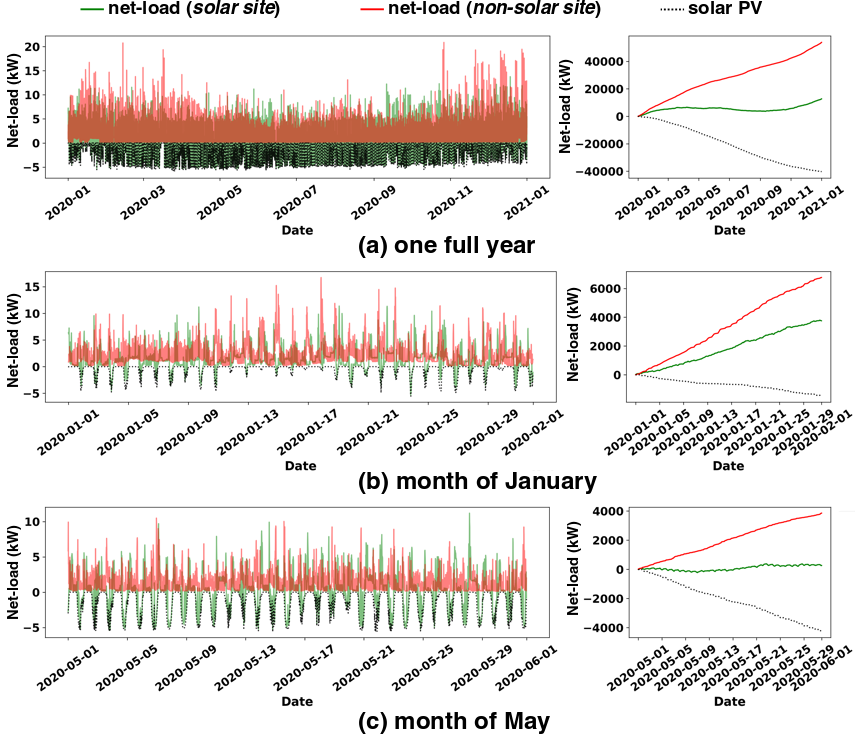}
    \caption{Comparison of net-load trends in the NEEA dataset for a \textit{solar} and \textit{a non-solar site} at various time-scales: (a) one full year, (b) a winter month (January), and (c) a spring/summer month (May). The plots on the right show the cumulative values of the data over the time period under consideration. The sub-metered solar PV data at the \textit{solar site} is also plotted.}
    \label{fig:neea_netload}
\end{figure}

Northwest Energy Efficiency Alliance (NEEA) has released a free publicly available dataset \cite{neeadataset} containing sub-metered time-series data of power consumption and generation from residential appliances and BTM solar PV units at 15-minute intervals. The data used in this work covers 28 months (September 2018 until December 2020) of power usage by 3380 circuits across 198 buildings. Additional weather-related measurements such as humidity and temperature (both interior and exterior) are also provided. In this work, we take the mean exterior temperature (if available) over all exterior temperature transmitters at each site. 
The data covers power consumption across several states and climatic zones. While the dataset does not specify the exact location (latitude and longitude) of each site, one may infer proximity of the test sites from the associated weather station IDs. The net-load of a site is defined as the measured power at the ‘Mains’ circuit of that site. Additionally, sites with solar installations have a dedicated ‘Solar PV’ circuit to report the generated solar PV power. 
%
%
The net-load profiles of a \textit{solar site} (i.e. a site with a solar installation) and a \textit{non-solar site} (i.e. a site without a solar installation) are illustrated in Fig.~\ref{fig:neea_netload}. The top row (Fig.\,\ref{fig:neea_netload}a) shows the data over one full year (Jan-Dec 2020), while the middle row (Fig.\,\ref{fig:neea_netload}b) and the bottom row (Fig.\,\ref{fig:neea_netload}c) display the data from the months of January (winter) and May (spring/summer), respectively. In each row, the right plot shows the cumulative values of the data over the time period under consideration. 
Due to the abundance of solar power in summer months, the net-load profile of the solar site is very close to zero. However, during winter months, solar production is meagre and the net-load of both solar and non-solar sites are somewhat comparable.
%
%
%
%
%
Furthermore, we identified different sites based on their solar penetration (solar generation divided by total energy consumption). We will evaluate our proposed forecasting model for various solar penetration for individual site, and also at an aggregated level across sites. 


\section{Description of the kPF-AE-LSTM Model}\label{sec3}

Our proposed deep learning model kPF-AE-LSTM has two key components: 1) a kernel-embedded Perron-Frobenius operated autoencoder (first introduced in \cite{huang2021forward}) to learn the latent space representation of the multi-dimensional input data; and 2) a probabilistic LSTM for generating the multi-timestep net-load forecast values. Before describing our proposed model and associated training process, it is important to introduce the readers to a few fundamental concepts associated with generative modeling, kPF operator, and probabilistic LSTM.

\subsection{Generative Modeling and Associated Training Complexity}
Given independent and identically distributed (i.i.d.) samples $X$ with an unknown distribution $P_X$, a generative model seeks to find a parametric distribution that closely resembles $P_X$. A popular example for a generative model is the variational autoencoder (VAE) \cite{kingma2016improved,huang2021forward} wherein the data generating distribution is learned via latent variables. In other words, we assume that there is some variable $Z\in\mathcal{Z}$ associated with the observed data $X\in\mathcal{X}$ that follows a known distribution $P_Z$ (also referred to as the prior in generative models). Then we can learn a mapping $f:\mathcal{Z}\rightarrow \mathcal{X}$ such that the distribution after transformation, denoted by $P_{f(Z)}$, aligns well with the data generating distribution $P_{X}$. Therefore, sampling from $P_{X}$ becomes convenient since $P_{Z}$ can be efficiently sampled. For VAEs, $f$ is parameterized by deep neural networks and $f$ is optimized using Maximum Likelihood Estimation (MLE), by minimizing the difference between $P_{f(Z)}$ and $P_{X}$. The operator which learns to minimize the distance between $P_{f(Z)}$ and $P_{X}$ is called a forward operator, as introduced in \cite{huang2021forward}.

This training process for finding the forward operator greatly influences the empirical performance of generative model. To learn the forward operator, VAEs use an approximate posterior $q_{Z|X}$ that may sometimes fail to align with the prior \cite{kingma2016improved}. This motivates the necessity of using kPF operator based process for training an AE model, as explained next, instead of training a VAE.

\subsection{Kernel-Embedded Perron-Frobenius (kPF) Operator}

Generative models such as VAE can be viewed as dynamical systems \cite{chen2018neural}, where the temporal evolution of the latent variable ($Z$) follows either a continuous or discrete time dynamics. Perron-Frobenius (PF) operator \cite{mayer1980ruelle} is a infinite dimensional linear operator, which can replace the previously mentioned forward operator such that $P_{X} =PF(P_{Z})$. $PF$ is infinite dimensional and there exists many finite dimensional estimation for $PF$. We have selected Kernel-embedded Perron-Frobenius (kPF) method to calculate finite dimensional estimation as introduced in \cite{klus2020eigendecompositions}.

\subsection{kPF-Autoencoder}

\begin{figure}
    \centering
    \includegraphics[width=\columnwidth]{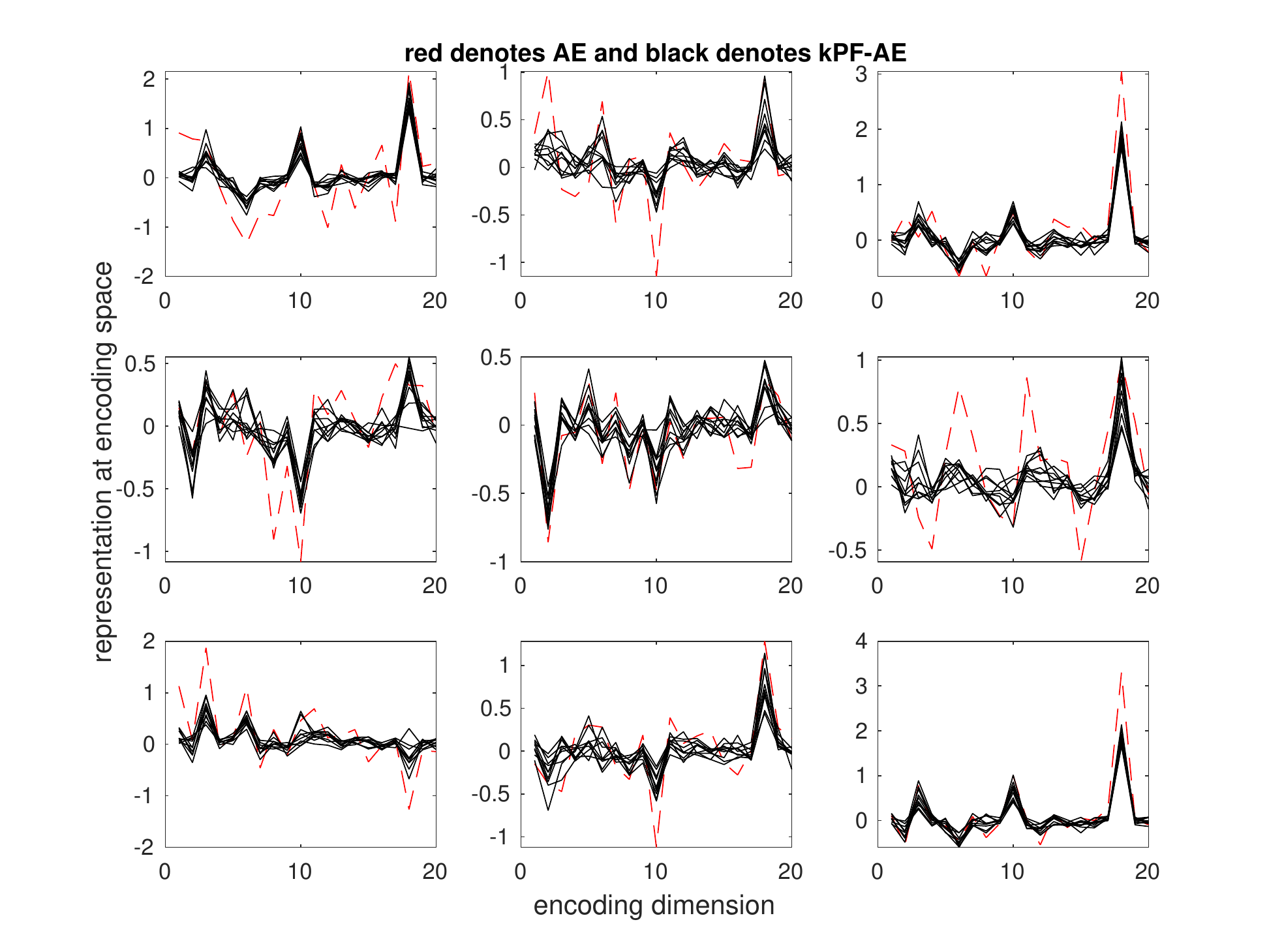}
    \caption{Samples from the calculated distributions across $20$ dimensional latent space. Different subplots represent the latent space distributions across different batches of the input dataset. The red plot denotes the latent representation after applying the convolutional AE. The black plot denotes the latent distributions after applying the proposed kPF-AE.}
    \label{fig:kPF}
\end{figure}

Our proposed kPF-Autoencoder (kPF-AE) architecture consists of a one-dimensional convolutional autoencoder, which finds a latent space (encoding space) to represent a historically (using temporal history) stacked input space. Specifically, this architecture is adapted from the one introduced in \cite{huang2021forward}, in which kPF-AE was used for two-dimensional image generation90, while we have modified the architecture for one-dimensional time-series forecasting. The kPF operator is applied on the resulting latent space using Algorithm\,\ref{alg1} (detailed derivation of which can be found in \cite{huang2021forward}) to generate the distribution of the latent variables. On applying the kPF-AE on the input space, we find the distribution of the input data in the latent space. A few of the samples from these latent space distributions are shown in Fig. \ref{fig:kPF}. We also would like to point out that Algorithm \ref{alg1} generates distribution across the whole latent space, calculated using the whole training dataset (irrespective of different batches of training data).

\begin{algorithm}
 \caption{Sample generation using kPF}
 \label{alg1}
 \begin{algorithmic}[1]
 \renewcommand{\algorithmicrequire}{\textbf{Input:}}
 \renewcommand{\algorithmicensure}{\textbf{Output:}}
 \REQUIRE trained encoder $\mathcal{E}$, training data $(\mathbf{x}_{1}, \mathbf{x}_{2},\dots, \mathbf{x}_{n})$, neighborhood size $\gamma$
 \ENSURE  generated $m$ samples $\mathbf{x}^{*}$ 
 \STATE Calculate encoded representation of training sample $\mathbf{X}_{e}=\Big(\mathcal{E}(\mathbf{x}_{1}),\mathcal{E}(\mathbf{x}_{2}),\dots,\mathcal{E}(\mathbf{x}_{n})\Big)$
 \STATE Calculate $K_{i,j}=k(\mathbf{X}_{e_{i}},\mathbf{X}_{e_{j}}),\forall(i,j)=1,\dots,n$, where $k(x,y):=\exp{\left(-{\|x-y\|}/{2}\right)}$
 \STATE Sample $z$ independently from $n$-variate standard Gaussian distribution $\mathcal{Z}$ 
\STATE Calculate $L_{i,j}=k(z_{i},z_{j}),\forall (i,j)= 1,\dots,n$
\STATE Calculate $K_{inv}=(K+n I)^{-1}$; $I$ is $n\times n$ identity matrix
\STATE Sample $w$ independently from $m$-variate standard Gaussian distribution $\mathcal{W}$
\STATE Calculate $V_{i,j}=k(z_{i},w_{j}),i\forall n,j\forall m$
\STATE Calculate $s=L\cdot K_{inv}\cdot V$
\STATE Calculate $\mathrm{ind}=\mathrm{argsort}(s)[-\gamma:]$
\STATE Calculate $\mathbf{x}^{*}={\mathbf{X}_{e}[\mathrm{ind}]s[\mathrm{ind}]} / {\|s[\mathrm{ind}]\|_{1}}$
 \end{algorithmic} 
 \end{algorithm}
 
 At this point, we shall introduce the concept behind probabilistic LSTM, and more specifically the concept of probabilistic prediction using recurrent neural networks.
 
 \subsection{Bayes by Back-Propagation (BBB)}
 Bayes by Back-propagation \cite{blundell2015weight} is a variational inference scheme for learning the posterior distribution on the weights $\theta\!=\!\lbrace\theta_i\rbrace_{i=1}^d\!\in\!\mathbb{R}^d$ of a neural network. The parameters $\theta$ are typically taken to be mutually independent Gaussian distributions, each with mean $\mu_i\!\in\!\mathbb{R}\,\forall i\!=\!1,\dots,d,$ and standard deviation $\sigma_i\!\in\!\mathbb{R}\,\forall i \!=\!1,\dots,d,$, denoted by $\mathcal{N}(\theta_i|\mu_i,\sigma_i^{2})$. Let $\log p(y|\theta,x)$ be the log-likelihood of the model, then the network is trained by minimizing the variational free energy, $\mathcal{L}(\theta)=\mathbb{E}_{q(\theta)}\Big[ \log \frac{q(\theta)}{p(y|\theta,x)p(\theta)}\Big]$\,.
 
 \subsection{BBB applied to Recurrent neural network (RNN)}
 The core of an RNN, $f$, is a neural network that maps the RNN state $s_{t}$ at step $t$, and an input observation $x_{t}$ to a new RNN state $s_{t+1},f:(s_{t},x_{t})\rightarrow s_{t+1}$.
 An RNN can be trained on a sequence of length $T$ by back-propagation through time by unrolling $T$ times into a feed-forward network. Explicitly, we set $s_t=f(s_{t-1},x_t),\forall t=1,\dots, T$. RNN parameters are learnt in much the same way as in a feed-forward neural network. A loss (typically after further layers) is applied to the states $s_{1:T}$ of the RNN, and then backpropagation is used to update the weights of the network. Crucially, the weights at each of the unrolled steps are shared. Thus each weight of the RNN core receives $T$ gradient contributions when the RNN is unrolled for $T$ steps. Applying BBB to RNNs, the weight matrices of the RNN are drawn from a distribution, whose parameters are learned. Therefore, the variational free energy for an RNN on a sequence of length $T$ is, $\mathcal{L}(\theta)=-\mathbb{E}_{q(\theta)}\Big[\log p(y_{1:T}|\theta,x_{1:T})\Big]+\mathrm{KL}\Big[q(\theta)||p(\theta)\Big]$.
 
\subsection{Proposed architecture}

\begin{figure*}
    \centering
    \includegraphics[scale=0.35]{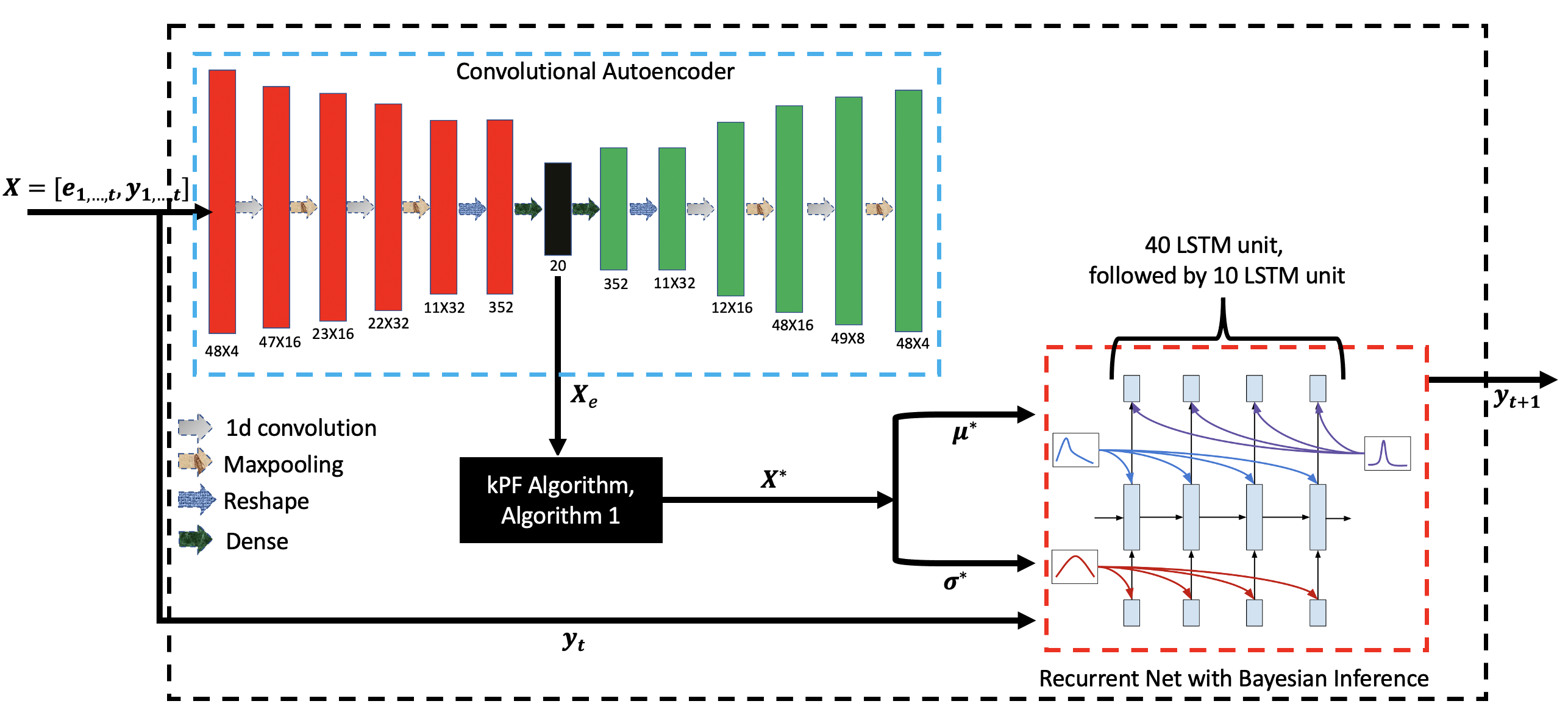}
    \caption{Schematic of our proposed architecture is shown within the black dashed box, which has primarily three sub-components, convolutional autoencoder (within blue dashed box), kPF Algorithm as shown in Algorithm \ref{alg1}, and a probabilistic LSTM network (within red dashed box).}
    \label{fig:arch}
\end{figure*}

Our proposed architecture is schematically shown in Fig. \ref{fig:arch}. Our framework takes temporal measurement consists of historic net-load ($y$) and weather related measurements ($e$) such as ambient temperature, relative humidity and wind direction. We will use the temporal stack of these variables to forecast net-load in future time. The main contribution of our architecture is the incorporation of the latent representation from the input space, in forecasting net-load. Although there exist methods in literature, which include historical input space information in the forecasting (such as use of Convolutional LSTM, LSTM etc.), the proposed model offers superior performance (as shown in next section) due to the incorporation of latent representation of the input space. Moreover, the proposed architecture also captures the variability present in the input dataset and is  able to forecast a distribution of net-load, unlike the existing methods. The proposed method also improves the training time via the incorporation of Algorithm \ref{alg1} over existing  methods such as variational recurrent neural network \cite{chung2015recurrent} for probabilistic forecasting. The details on training time improvement is given in the next section.

\section{Results and Discussion}\label{sec4}

\subsection{Data-Splitting}\label{splitting}
In order to evaluate the performance of our proposed architecture, we segment the data into training and test sets involves allocating the first $f\%$ (say $f=75$) as training and the rest as test. This kind of data-partitioning will be admissible if the statistics of the training data are not too different from that of the test data. However, in our applications, the statistics of solar PV or net-load are highly dependent on the season. Therefore, such a mode of partition may result in a highly-biased model, which is trained mostly on, say, summer data.
\begin{figure}
    \centering
    \includegraphics[width=0.48\textwidth]{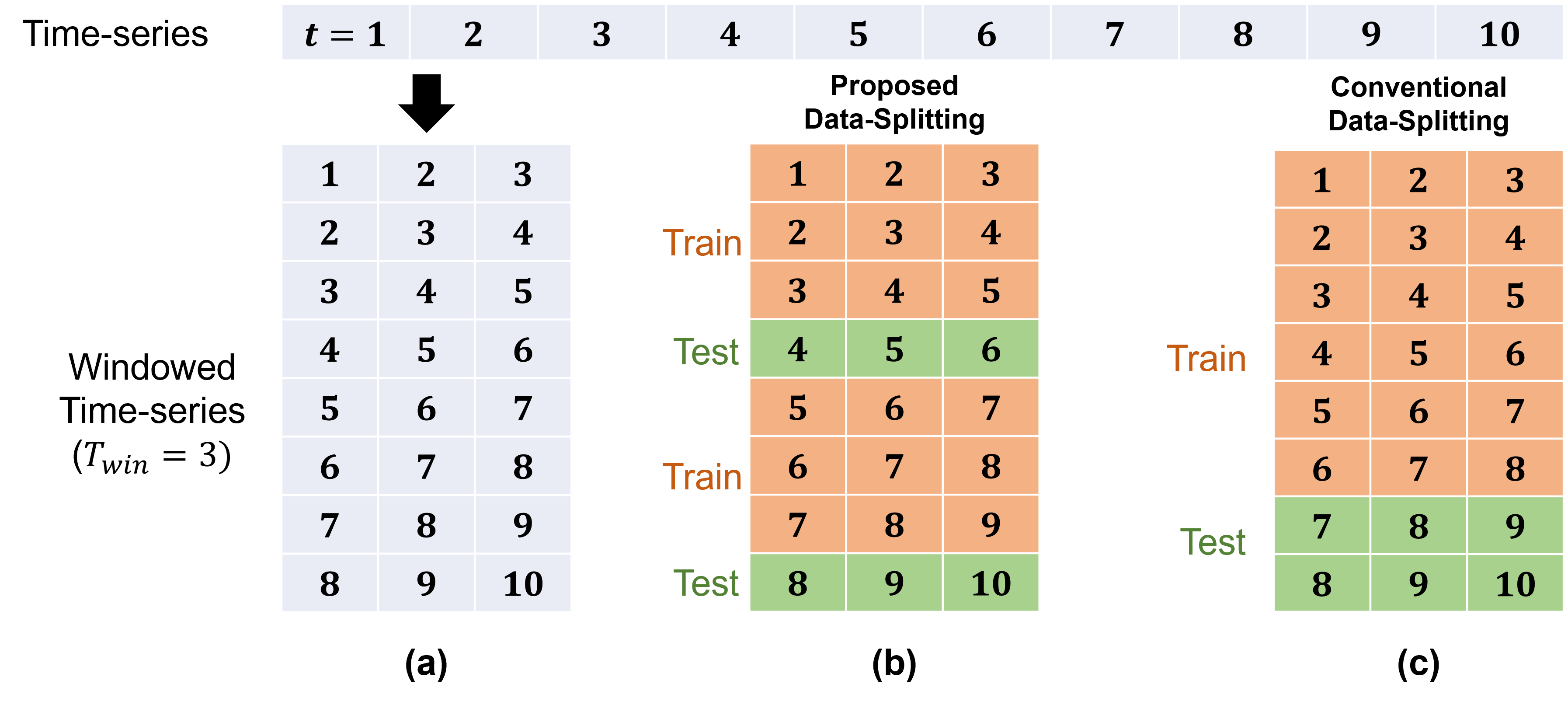}
    \caption{Illustration of  (a) time-series windowing and (b) proposed data-splitting scheme and (c) conventional data-splitting scheme. Training using the proposed scheme results in lesser bias due to seasonality of the data.}
    \label{fig:data-splitting}
\end{figure}
To circumvent this problem, we first 'window' the data as usual, based on the window-size ($T_{win}$) chosen as input to the model (Fig.~\ref{fig:data-splitting}a). Then we allocate the first 3 (say) windows as training, the next 1 (say) window as test, and so on for the entire dataset (Fig.~\ref{fig:data-splitting}b). In this way, we obtain an approximate 75/25 split, which is temporally unbiased. The conventional data-splitting method for obtaining a similar split, is also shown in Fig.~\ref{fig:data-splitting}c, wherein the first 6 windows are tagged as training and the last 2 as testing.

\subsection{Metrics}\label{metrics}
Several metrics are used to test and compare the performance and generalizability of various models. 
Commonly used metrics for deterministic forecasts include the mean absolute error (MAE), and the mean absolute percentage error (MAPE). 
%
Additionally, we also consider a couple of additional performance metrics which are more appropriate for probabilistic forecast models. 
One such metric is the continuous ranked probability score (CRPS) \cite{van2018review,gneiting}. Given a predicted cumulative density function (CDF), $F_{pred}(y)$\,, and an observed value, $y_{obs}$\,, CRPS measures the cumulative quadratic discrepancy between $F_{pred}(y)$ and x-axis for all $y<y_{obs}$, and that between $F(y)=1$ and $F_{pred}(y)$ for all $y\ge y_{obs}$. Mathematically, the CRPS is defined as:
%
\begin{align}
    &crps(F_{pred},y_{obs}) = \alpha+\beta\,,\quad \text{where,}\label{eq:crps_func_concise}\\
    &\alpha \!:=\! \int_{-\infty}^{y_{obs}}\!\left(F_{pred}(y)\right)^2 dy\,,\,~\beta \!:=\! \int_{y_{obs}}^{\infty}\!\left(1\!-\!F_{pred}(y)\right)^2 dy\,.\notag
\end{align}
\begin{figure}
    \centering
    \includegraphics[width=0.45\textwidth]{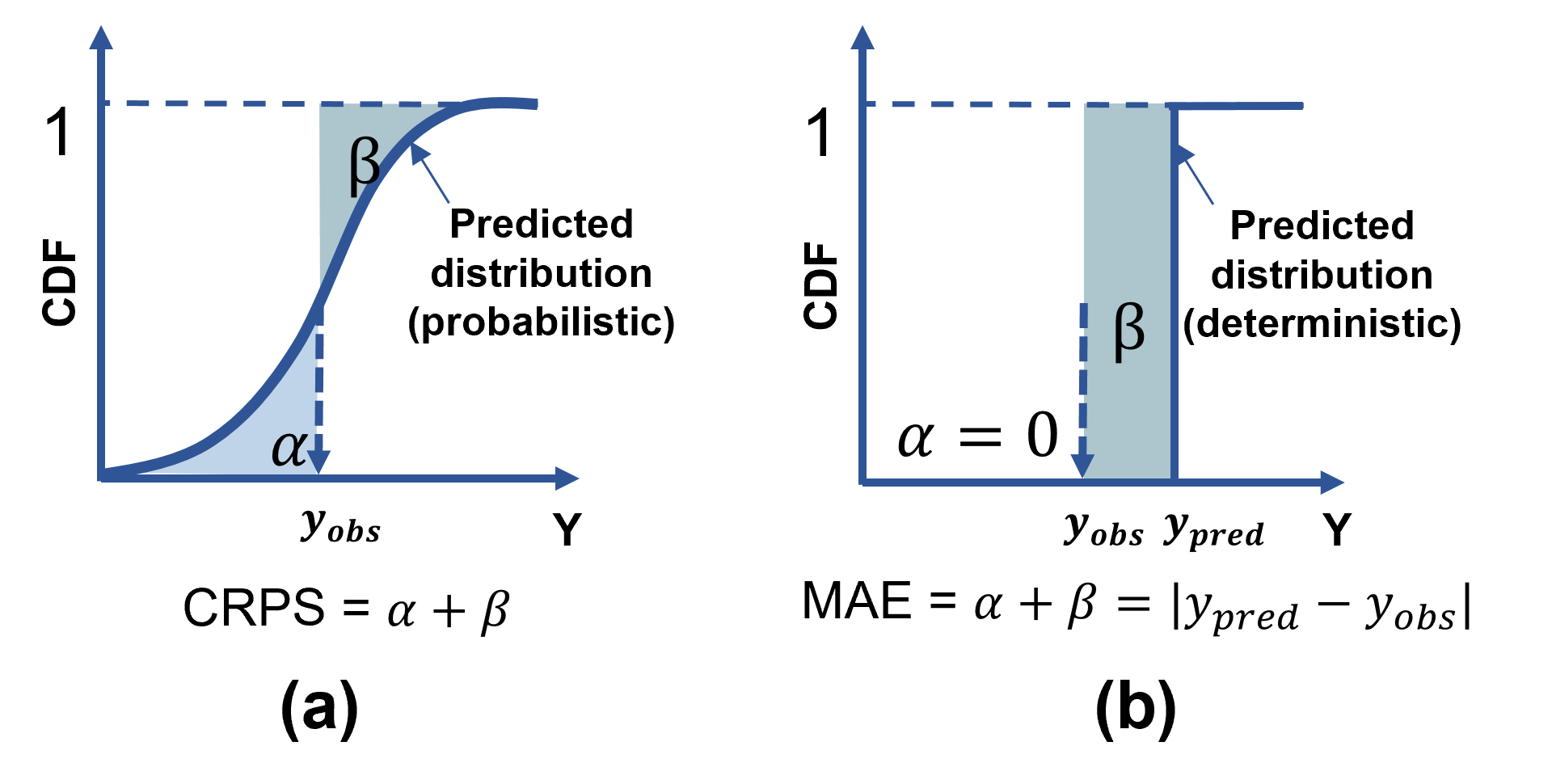}
    \caption{Illustration of (a) CRPS and (b) MAE. CRPS is a metric function used when the predicted output is a probability distribution. Mathematically, CRPS simplifies into the MAE when the output is deterministic.}
    \label{fig:crps_vs_mae}
\end{figure}
This is also shown in Fig.~\ref{fig:crps_vs_mae}(a), where $\alpha,\beta$ are the quadratic discrepancies for the two section of  $F_{pred}(y)$. It is clear from the Fig.~\ref{fig:crps_vs_mae}(b) that, when the predicted value is deterministic, wherein $F_{pred}(y)$ takes the form of a step function, the CRPS function simplifies into the MAE. Moreover, if we assume the predicted probability distributions to be Gaussian with mean $\mu_{pred}$ and variance $\sigma_{pred}^2$, then the $crps(\cdot)$ assumes a closed-form expression \cite{gneiting}.
As a metric, we compute the average CRPS across the entire training dataset as 
\begin{align}
CRPS = \frac{1}{T}\sum_{t=1}^{T}crps(F_{pred}(t),y_{obs}(t))
\label{eq:crps_avg}
\end{align}
Another alternative metric for probabilistic outputs is the \textit{probability between bounds} (PBB), defined as: 
\begin{equation}
    PBB = P\left(\underline{y}_{pred}<y_{obs}<\overline{y}_{pred}\right)
    \label{eq:pbb}
\end{equation}
which is the fraction of observations lying between the predicted upper ($\overline{y}_{pred}$) and lower ($\underline{y}_{pred}$) bounds from the probabilistic model. 
In this work, we select $\overline{y}_{pred} \!=\! \mu_{pred}\!+\!\sigma_{pred}$ and $\underline{y}_{pred} \!=\! \mu_{pred}\!-\!\sigma_{pred}$\,,
where $\mu_{pred}$ and $\sigma_{pred}$ denote the predicted mean and standard deviation, respectively.


\subsection {Performance Analysis}
We evaluate the robustness of the forecast performance of the proposed kPF-AE-LSTM model under various scenarios, including different behind-the-meter solar penetration levels (up to 50\%) and different levels of aggregation of sites, as well as compare the model with existing deep forecasting models. After splitting the NEEA dataset as mentioned in Sec.\,\ref{splitting}, we augment the train and test split, by temporarily stacking all the input variables, such as the ambient temperature, relative humidity, and apparent power. For 24 hours ahead forecast, 480 preceding time-points (i.e.,  5 days at 15 minutes resolution) of measurements were temporally stacked, while 48 preceding time-points (i.e.,  12 hrs) of measurements for 15 min ahead forecast. Recall from Sec.\,\ref{sec3} that the model generates the predicted mean ($\mu_{pred}(t)$) and the standard deviation ($\sigma_{pred}(t)$) as a time-series output. Therefore, we can use the model output to construct various confidence intervals around the prediction bounds, e.g., $\mu_{pred}\!\pm\!2\sigma_{pred}$ generating $\sim$95\% confidence intervals for Gaussian distributed output.

\subsubsection{Illustrative Results} 
\begin{figure}
    \centering
    \includegraphics[width=\columnwidth]{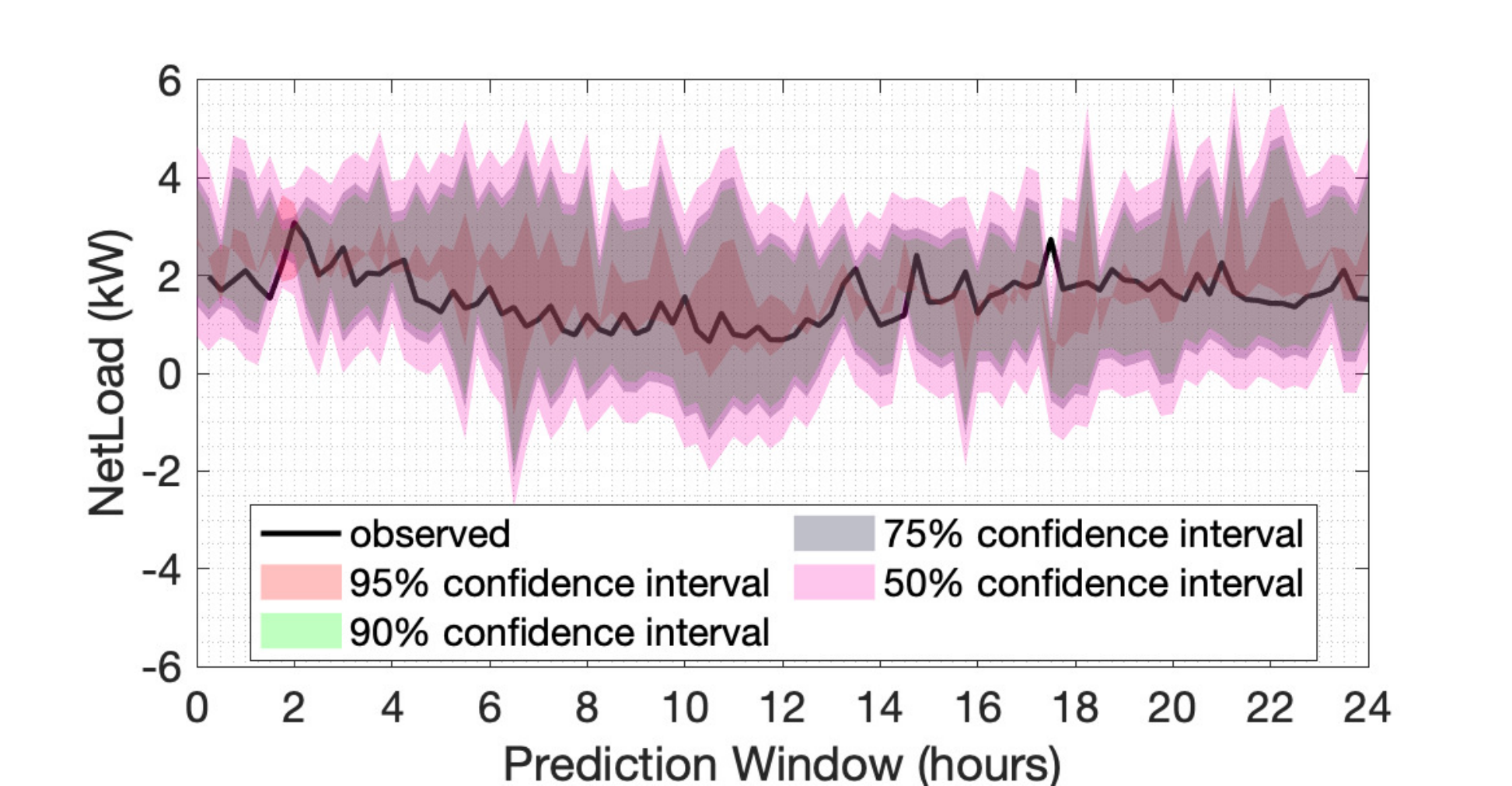}
    \caption{Probabilistic forecast for a site with $50\%$ solar penetration. We have picked $24$ consecutive hours to plot the forecast from our proposed model across different confidence intervals. The observed (measured) net-load value is also plotted.}
    \label{fig:corr1}
\end{figure}
We illustrate in Fig. \ref{fig:corr1} the performance of our proposed model across a 24 hours prediction window for an individual site with 50\% behind-the-meter solar penetration in terms of annual energy. Different colored regions in the plot indicate different confidence intervals of the forecast and the `black' line denotes the observed net-load value across this 24 hour window. Overall, with 95\% confidence interval, our model is able to capture 99\% of the observed data-points, while it captures 75\% of the observed data-points with 50\% confidence interval. 

\begin{figure}[!ht]

\subfloat[]{%
  \includegraphics[width=\columnwidth]{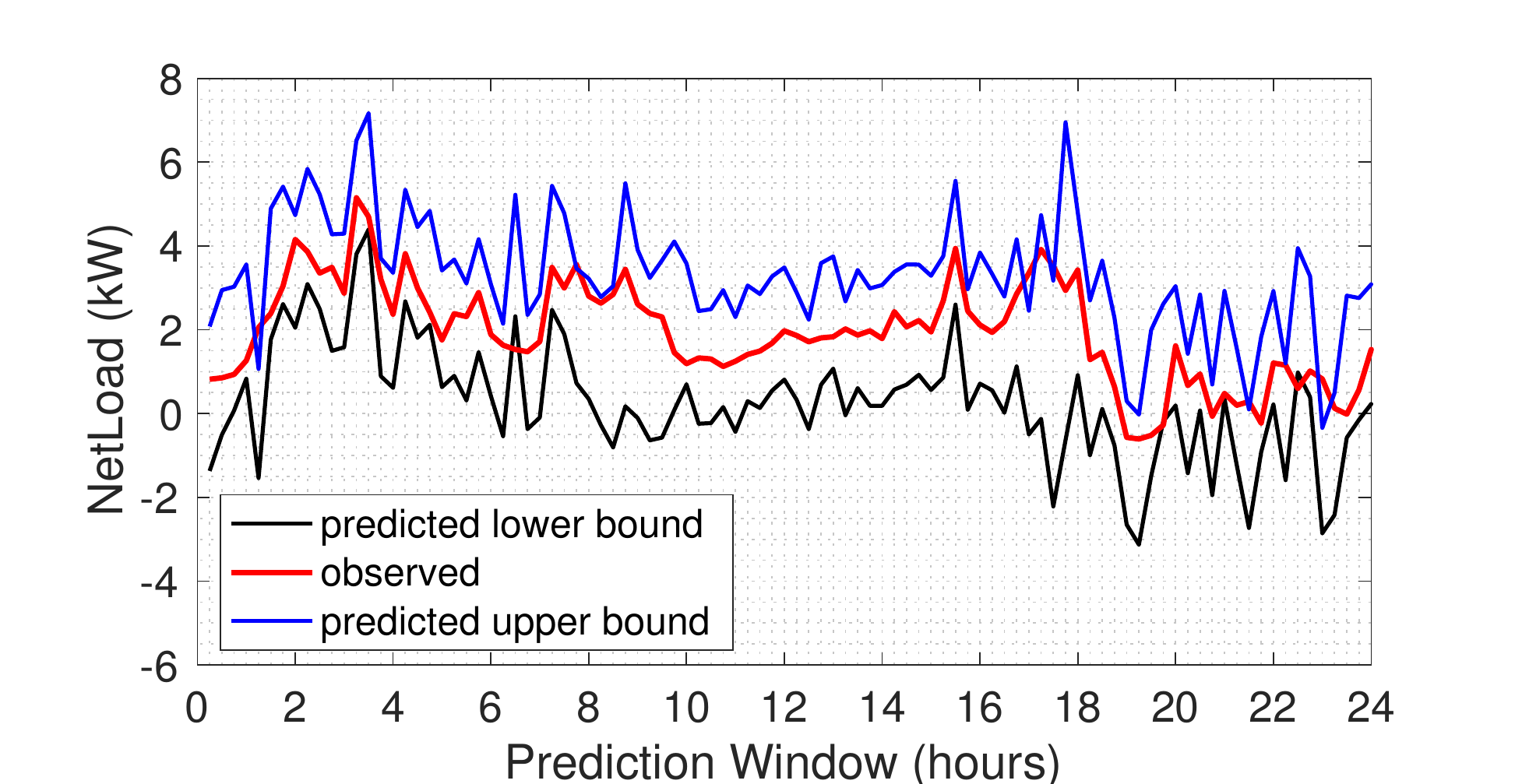}%
}

\vspace{-0.1in}

\subfloat[]{%
  \includegraphics[width=\columnwidth]{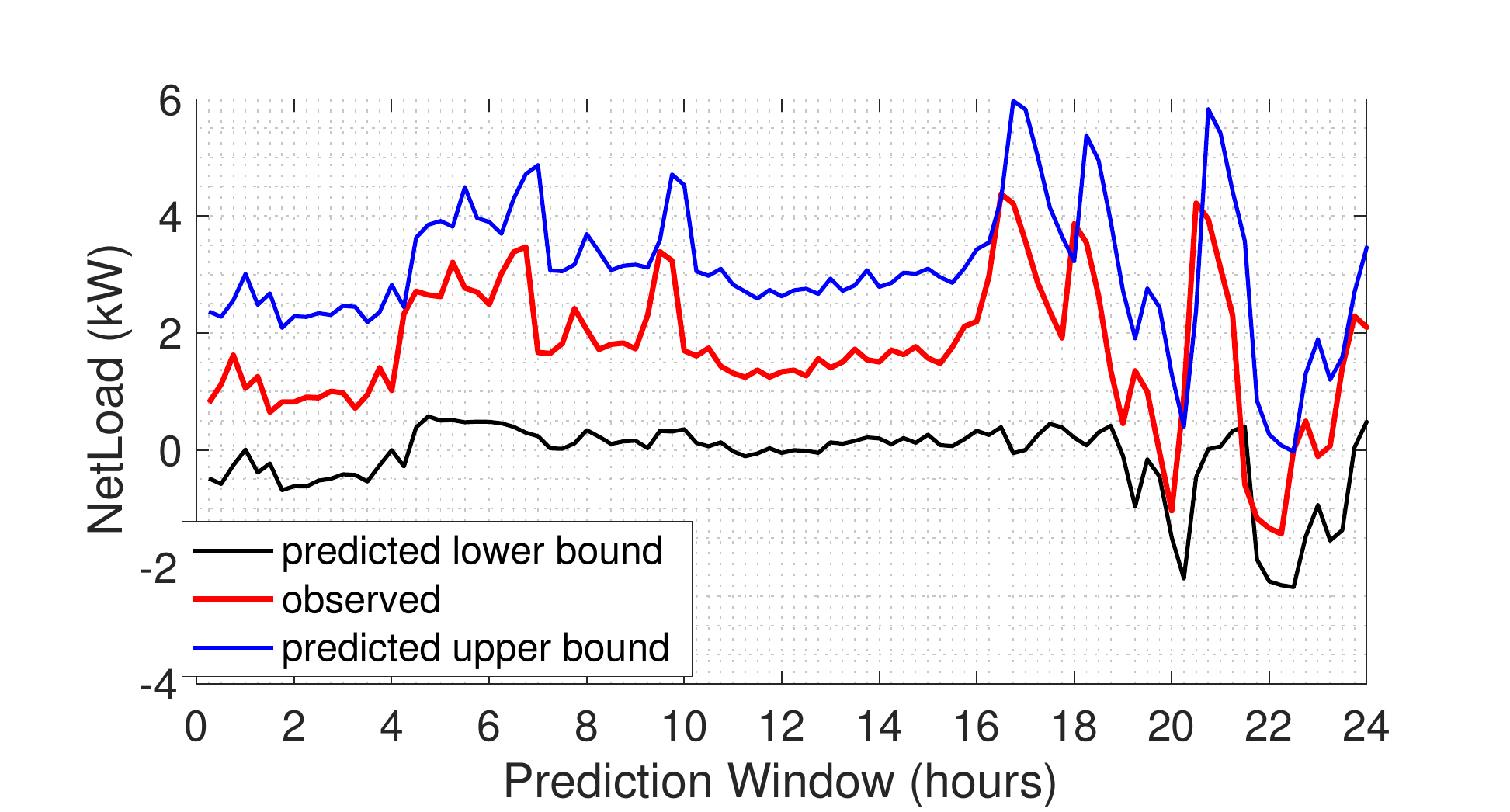}%
}

\caption{Prediction bounds with 95\% confidence is plotted across the observed net-load value (in green), for a site with 50\% solar penetration, for two different 24 hours window.}\label{fig1}

\end{figure}

\begin{figure}[!ht]

\subfloat[]{%
  \includegraphics[width=\columnwidth]{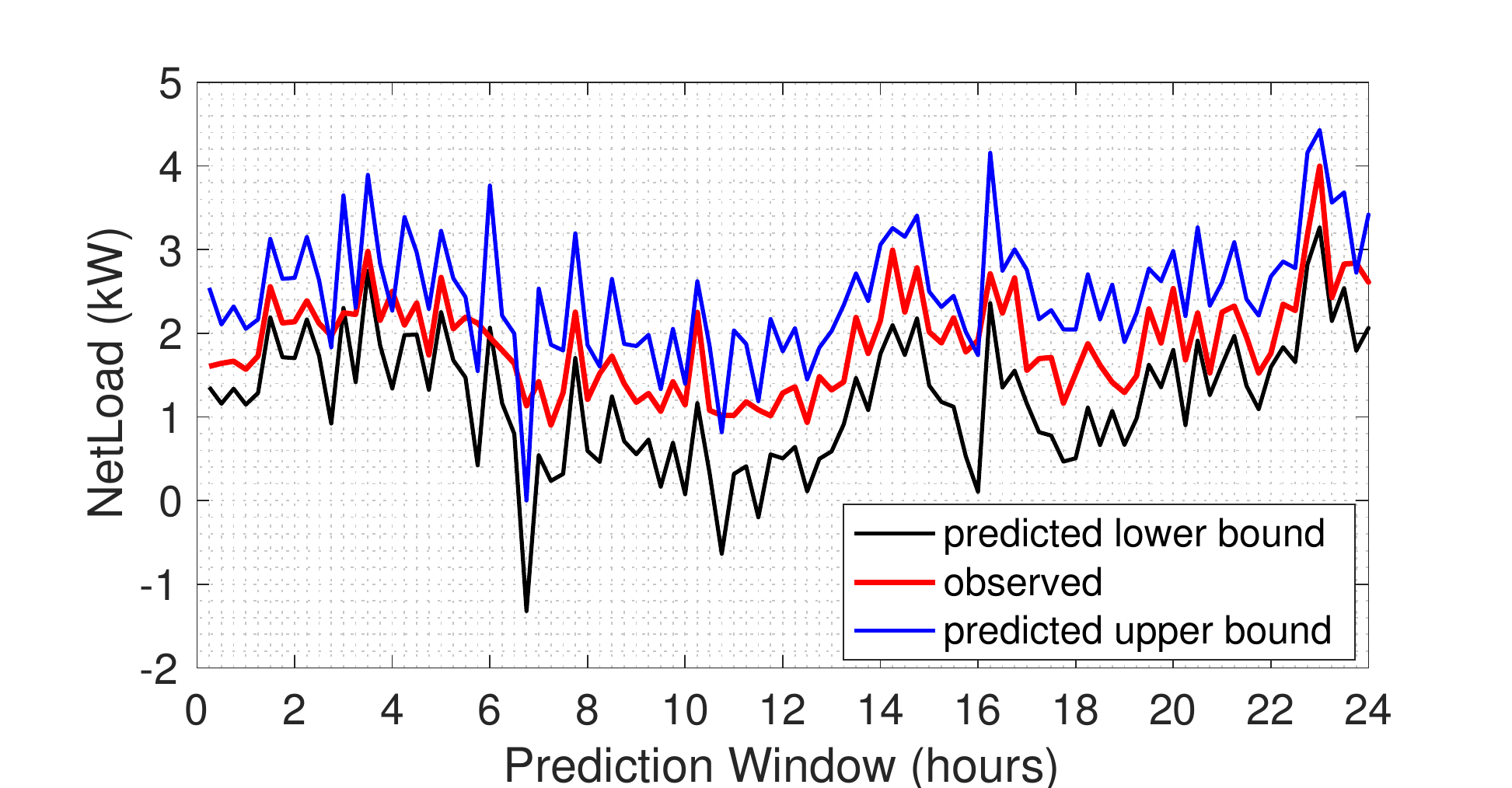}%
}

\vspace{-0.1in}

\subfloat[]{%
  \includegraphics[width=\columnwidth]{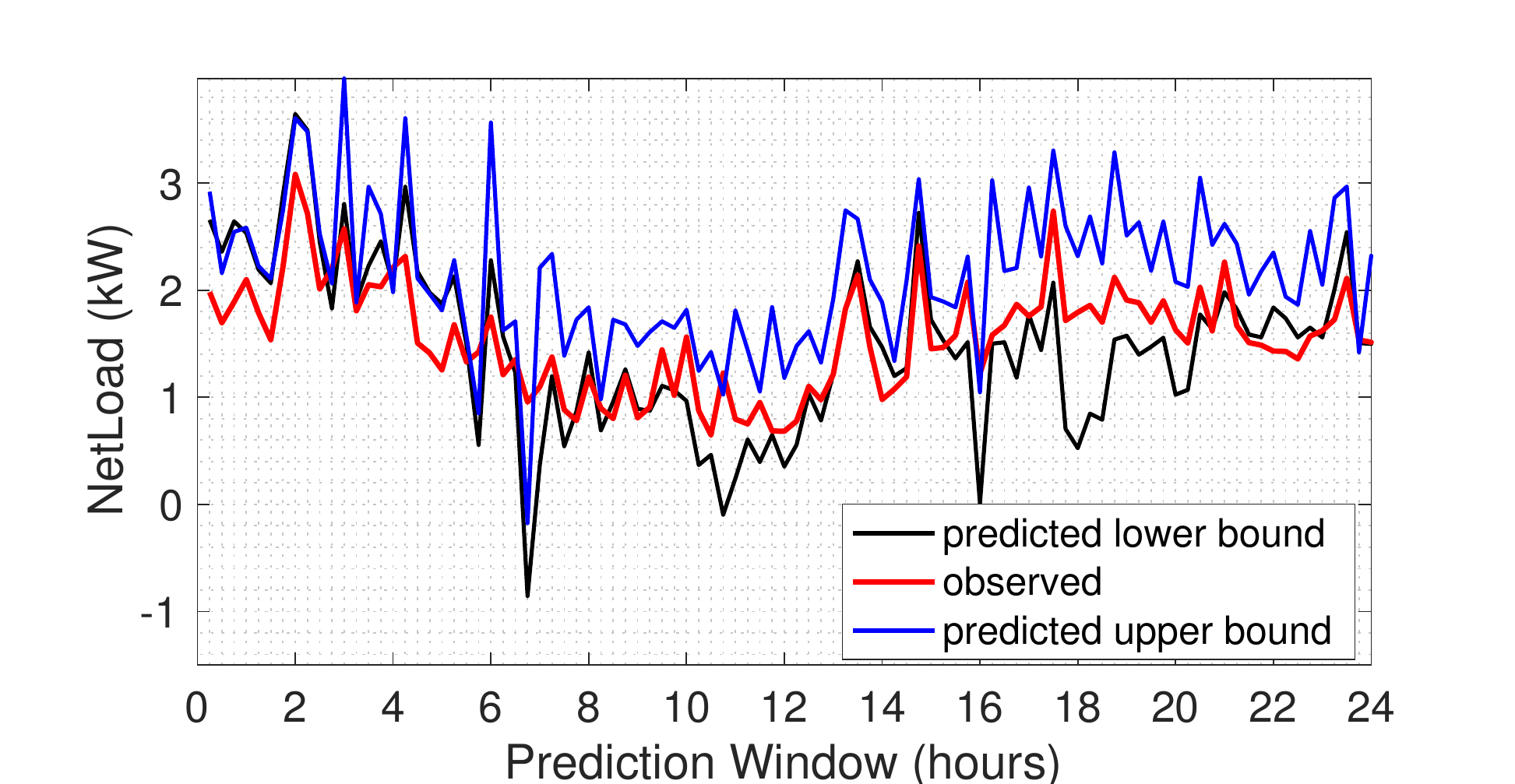}%
}

\caption{Prediction bounds with 95\% confidence is plotted across the observed net-load value (in green), for a site with no solar, for two different 24 hours window.}\label{fig2}
\end{figure}

In Figs.\,\ref{fig1} and \ref{fig2}, we have plotted the 95\% confidence intervals of 24\,hrs ahead forecast from the test set for two different sites: one with 50\% solar penetration and the other with no solar, respectively. Note the relatively larger width of the 95\% confidence interval in 50\% solar site, as compared to the non-solar site, which reflects the larger variability in corresponding net-load due to solar. We also noticed that the observed data points sometimes lie outside of the 95\% confidence interval at the non-solar site (Fig.\,\ref{fig2}) which may be attributed to relatively lower data availability for the non-solar site -- approximately 20\% less training data for the non-solar site compared to the 50\% solar site. 

\subsubsection{Latent Space Representation} 
\begin{figure}
    \centering
    \includegraphics[width=\columnwidth]{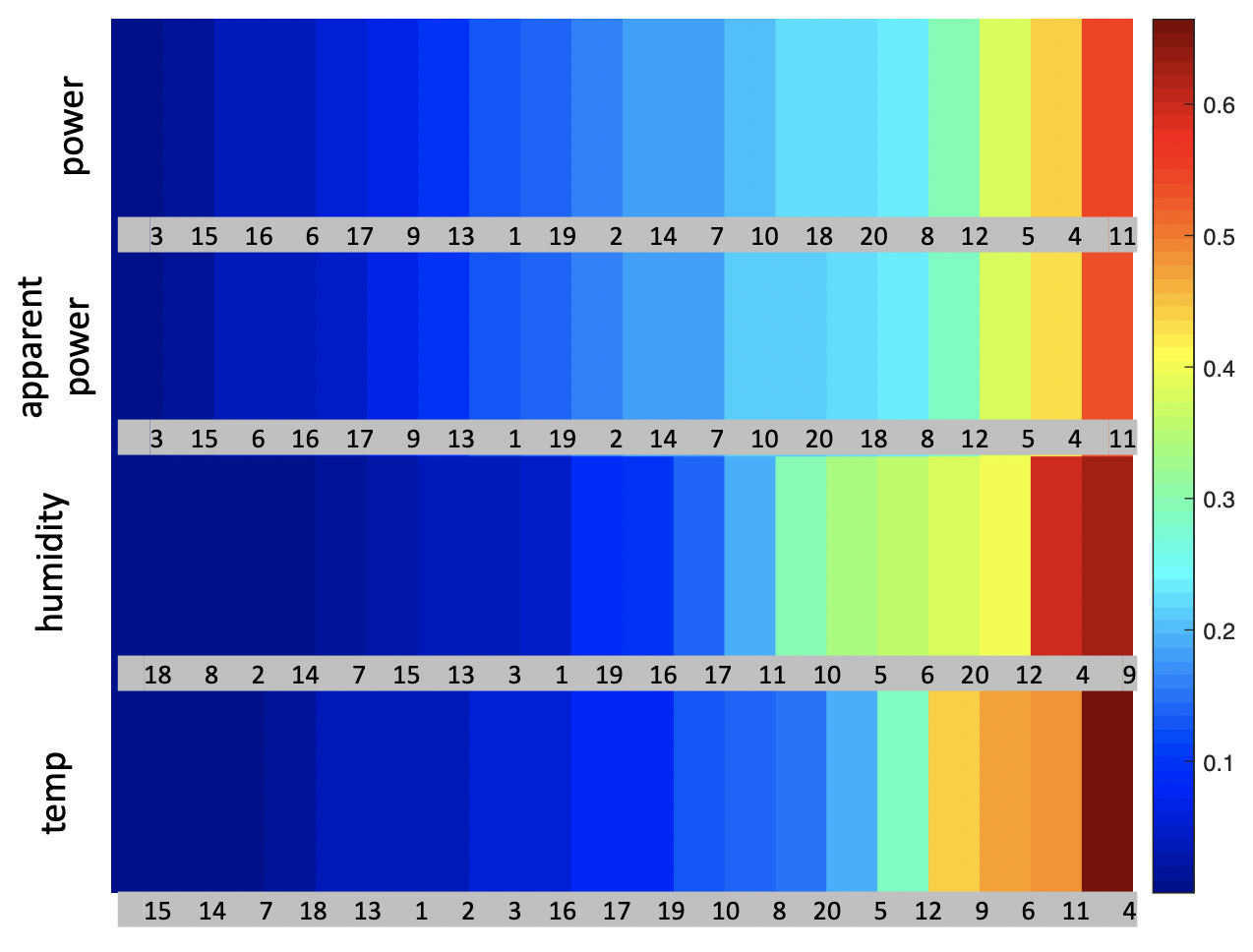}
    \caption{Heatmap of the correlation matrix between input space and the latent space, calculated using the convolutional autoencoder. The correlation shows relatively more number of variables in latent space, representative of ambient temperature compared to other variables.}
    \label{fig:corr}
\end{figure}
We illustrate how the input features are statistically encoded within the latent (encoding) space which is learnt via the sequential training process described in Sec.\,\ref{sec3}. In Fig. \ref{fig:corr}, we plot the correlation matrix between the 20-dimensional encoding/latent space and the input variables. The heatmap shows that ambient temperature, within all the input variables, got represented the most in the latent space, with four variables in latent space (namely, \#4,\,11,\,6,\,and\,9) having correlation larger than 0.45). Moreover, the latent space variable \#11 seems to be strongly correlated with the (both active and apparent) power input, while the latent variables \#9 and 4 are strongly correlated with humidity.

\subsubsection{Benchmarking at Varying Solar Penetration}
We evaluate and benchmark the proposed model performance against existing probabilistic forecast models, under varying levels of uncertainty introduced due to increasing behind-the-meter solar penetration. Focusing on a particular weather station (weather site-id 720638-224), we have computed solar penetration of different (individual) sites in terms of annual energy, and selected five sites with solar penetrations of: 0\%, 10\%, 20\%, 36\%, and 50\%. We have trained and tested our proposed model separately on each of these selected sites. 
%
%
\begin{table*}[]
\centering
\begin{tabular}{|c|c|c|c|c|c|c|c|c|c|c|c|}
\hline
\multirow{5}{*}{\rotatebox[origin=c]{90}{\textbf{\begin{tabular}[c]{@{}c@{}} Prediction\\ Window\end{tabular}}}} & \multirow{5}{*}{\rotatebox[origin=c]{90}{\textbf{Method}}} & \multirow{5}{*}{\rotatebox[origin=c]{90}{\textbf{\begin{tabular}[c]{@{}c@{}}Solar \\ Penetration\\ (\%)\end{tabular}}}} & \multirow{5}{*}{\textbf{\begin{tabular}[c]{@{}c@{}}Overall\\ MAE\\ (kW)\end{tabular}}} & \multicolumn{4}{c|}{} & \multicolumn{4}{c|}{}\\
& & & & \multicolumn{4}{c|}{\textbf{Winter}} & \multicolumn{4}{c|}{\textbf{Summer}} \\ 
& & & & \multicolumn{4}{c|}{} & \multicolumn{4}{c|}{}\\
\cline{5-12} 
& &  &  & {\textbf{MAE}} & {\textbf{MAPE}} & {\textbf{PBB}} & \textbf{CRPS} & {\textbf{MAE}} & {\textbf{MAPE}} & {\textbf{PBB}} & \textbf{CRPS} \\
 & & & & {\textbf{(kW)}} & {\textbf{(\%)}} & {\textbf{(\%)}} & & {\textbf{(kW)}} & {\textbf{(\%)}} & {\textbf{(\%)}} &\\
 \hline\hline
\multirow{10}{*}{\rotatebox[origin=c]{90}{15 minutes}} & \multirow{5}{*}{\rotatebox[origin=c]{90}{pCLSTM}}   & 0 & 0.72 & {0.85} & {9.20} & {64.36} & 0.38 & {0.59} & {9.14} & {68.74} & 0.40 \\ \cline{3-12}
 && 10 & 0.75 & {0.90} & {9.41} & {61.38} & 0.38 & {0.60} & {9.82} & {65.66} & 0.44 \\ \cline{3-12} 
 && 20 & 0.86 & {0.99} & {10.13} & {64.76} & 0.36 & {0.73} & {10.15} & {69.87} & 0.33 \\ \cline{3-12}
 && 36 & 1.18 & {1.30} & {10.10} & {65.87} & 0.33 & {1.05} & {10.27} & {65.78} & 0.37 \\ \cline{3-12} 
 && 50 & 1.55 & {1.60} & {11.20} & {72.24} & 0.29 & {1.51} & {11.21} & {67.50} & 0.39 \\ 
 \cline{2-12}
& \multirow{5}{*}{\rotatebox[origin=c]{90}{\textit{\textbf{\begin{tabular}[c]{@{}c@{}}kPF-\\AE-LSTM\end{tabular}}}}} & \textit{\textbf{0}} & \textit{\textbf{0.51}} & {\textit{\textbf{0.56}}} & {\textbf{4.13}} & {\textit{\textbf{91.18}}} & \textit{\textbf{0.09}} & {\textit{\textbf{0.34}}} & {\textbf{4.12}} & {\textit{\textbf{92.23}}} & \textit{\textbf{0.10}} \\ \cline{3-12} 
&& \textit{\textbf{10}} & \textit{\textbf{0.51}} & {\textit{\textbf{0.62}}} & {\textbf{4.10}} & {\textit{\textbf{89.80}}} & \textit{\textbf{0.10}} & {\textit{\textbf{0.40}}} & {\textbf{4.19}} & {\textit{\textbf{91.14}}} & \textit{\textbf{0.09}} \\ \cline{3-12} 
&& \textit{\textbf{20}} & \textit{\textbf{0.60}} & {\textit{\textbf{0.72}}} & {\textbf{4.22}} & {\textit{\textbf{92.29}}} & \textit{\textbf{0.09}} & {\textit{\textbf{0.49}}} & {\textbf{4.29}} & {\textit{\textbf{93.34}}} & \textit{\textbf{0.11}} \\ \cline{3-12} 
&& \textit{\textbf{36}} & \textit{\textbf{0.82}} & {\textit{\textbf{0.56}}} & {\textbf{4.31}} & {\textit{\textbf{92.23}}} & \textit{\textbf{0.08}} & {\textit{\textbf{0.70}}} & {\textbf{4.36}} & {\textit{\textbf{90.31}}} & \textit{\textbf{0.11}} \\ \cline{3-12} 
&& \textit{\textbf{50}} & \textit{\textbf{1.08}} & {\textit{\textbf{0.74}}} & {\textbf{4.56}} & {\textit{\textbf{94.51}}} & \textit{\textbf{0.09}} & {\textit{\textbf{0.80}}} & {\textbf{4.87}} & {\textit{\textbf{92.23}}} & \textit{\textbf{0.11}} \\ 
\hline\hline
\multirow{10}{*}{\rotatebox[origin=c]{90}{24 hours}} &\multirow{5}{*}{\rotatebox[origin=c]{90}{pCLSTM}}  & 0 & 1.12 & {1.32} & {14.31} & {59.20} & 0.42 & {0.92} & {14.22} & {58.17} & 0.43 \\ \cline{3-12}
 && 10 & 1.24 & {1.49} & {15.55} & {60.81} & 0.43 & {0.99} & {16.23} & {59.98} & 0.45 \\ \cline{3-12}
 && 20 & 1.31 & {1.51} & {15.43} & {61.19} & 0.40 & {1.11} & {15.46} & {60.87} & 0.41 \\ \cline{3-12}
 && 36 & 1.29 & {1.42} & {11.04} & {62.28} & 0.38 & {1.14} & {11.22} & {61.18} & 0.40 \\ \cline{3-12}
 && 50 & 1.45 & {1.50} & {10.47} & {67.71} & 0.35 & {1.41} & {10.48} & {62.23} & 0.38 \\ 
 \cline{2-12}
&\multirow{5}{*}{\rotatebox[origin=c]{90}{\textit{\textbf{\begin{tabular}[c]{@{}c@{}}kPF-\\AE-LSTM\end{tabular}}}}} & \textit{\textbf{0}} & \textit{\textbf{0.71}} & {\textit{\textbf{0.78}}} & {\textbf{5.75}} & {\textit{\textbf{87.71}}} & \textit{\textbf{0.14}} & {\textit{\textbf{0.47}}} & {\textbf{5.73}} & {\textit{\textbf{86.67}}} & \textit{\textbf{0.16}} \\ 
\cline{3-12}
&& \textit{\textbf{10}} & \textit{\textbf{0.76}} & {\textit{\textbf{0.92}}} & {\textbf{6.11}} & {\textit{\textbf{86.18}}} & \textit{\textbf{0.16}} & {\textit{\textbf{0.59}}} & {\textbf{6.24}} & {\textit{\textbf{85.51}}} & \textit{\textbf{0.17}} \\ 
\cline{3-12}
&& \textit{\textbf{20}} & \textit{\textbf{0.82}} & {\textit{\textbf{0.98}}} & {\textbf{5.77}} & {\textit{\textbf{88.52}}} & \textit{\textbf{0.15}} & {\textit{\textbf{0.68}}} & {\textbf{5.86}} & {\textit{\textbf{87.78}}} & \textit{\textbf{0.17}} \\ 
\cline{3-12}
&& \textit{\textbf{36}} & \textit{\textbf{0.91}} & {\textit{\textbf{0.62}}} & {\textbf{4.78}} & {\textit{\textbf{84.45}}} & \textit{\textbf{0.14}} & {\textit{\textbf{0.78}}} & {\textbf{4.83}} & {\textit{\textbf{83.31}}} & \textit{\textbf{0.16}} \\ 
\cline{3-12}
&& \textit{\textbf{50}} & \textit{\textbf{1.14}} & {\textit{\textbf{0.78}}} & {\textbf{4.81}} & {\textit{\textbf{85.57}}} & \textit{\textbf{0.15}} & {\textit{\textbf{0.84}}} & {\textbf{5.14}} & {\textit{\textbf{82.23}}} & \textit{\textbf{0.16}} \\ 
\hline
\end{tabular}
\caption{Comparison of forecast performance at various prediction windows (15-min and 24-hr ahead) -- measured via different metrics -- between our proposed model (kPF-AE-LSTM) and a probabilistic convolutional LSTM (pCLSTM) model, for individual sites with various solar penetration within the NEAA dataset.}\label{table11}
\end{table*}
We compared our model performance with existing probabilistic methods, such as probabilistic convolutional-LSTM (pCLSTM)\cite{wang2019probabilistic}. The results in Table\,\ref{table11} demonstrate the superiority of our model over the selected baseline in all metrics, mentioned in Section \ref{metrics}, for 15\,min and day-ahead forecasts. Specifically, there is a $30\%$ improvement in probabilistic performance metric, such as PBB and CRPS, across five different solar penetration levels (0\% to 50\%). We have not found any significant performance difference of our model between summer and winter season.

\begin{table}[]
\centering
\begin{tabular}{|c|c|c|ccc|}
\hline
\multirow{3}{*}{\textbf{Method}} & \multirow{3}{*}{\textbf{\begin{tabular}[c]{@{}c@{}}Weights\\ ($\times 1000$)\end{tabular}}} & \multirow{3}{*}{\textbf{\begin{tabular}[c]{@{}c@{}}Train\\ Time\\ (s)\end{tabular}}} & \multicolumn{3}{c|}{\textbf{\begin{tabular}[c]{@{}c@{}}Forecast Performance\end{tabular}}} \\ \cline{4-6} 
 &  &  & \multicolumn{1}{c|}{\bfseries MAE} & \multicolumn{1}{l|}{\bfseries PBB} & {\bfseries CRPS} \\ 
  &  &  & \multicolumn{1}{c|}{\bfseries (kW)} & \multicolumn{1}{l|}{\bfseries (\%)} & \\\hline
pCLSTM & 47 & \multicolumn{1}{c|}{6.0} & \multicolumn{1}{c|}{1.34} & \multicolumn{1}{l|}{62.27} & 0.41 \\ \hline
VRNN & 43 & 9.3 & \multicolumn{1}{c|}{0.85} & \multicolumn{1}{l|}{84.13} & 0.16 \\ \hline
\textit{\textbf{kPF-AE-LSTM}} & \textit{\textbf{25}} & \textit{\textbf{2.0}} & \multicolumn{1}{c|}{\textit{\textbf{0.86}}} & \multicolumn{1}{l|}{\textit{\textbf{86.27}}} & \textit{\textbf{0.17}} \\ \hline
\end{tabular}
\caption{Performance comparison between our proposed model, variational recurrent neural network (VRNN) \cite{chung2015recurrent}, and pCLSTM, for a day-ahead forecast.}\label{table2}
\end{table}
In addition to the improvement in performance metrics, our proposed model is shown to be $\sim$80\% computationally more efficient compared to the closest performing variational recurrent neural network (VRNN), as documented in Table \ref{table2}. Our model also consumes $\sim$50\% less computational memory when compared to VRNN. The VRNN, although architecturally similar to our proposed model, our model achieves faster training convergence due to the usage of Algorithm \ref{alg1}. VRNN consists of simultaneous minimization of several loss functions (reconstruction, divergence, and regression), which takes significantly more training epoch for convergence. This further shows the benefit of our proposed model in terms of both performance as well as ease of training.


\subsubsection{Varying Level of Aggregation}
\begin{table*}[]
\centering
\begin{tabular}{|c|c|c|clcc|clcc|}
\hline
\multirow{2}{*}{\textbf{Method}} & \multirow{2}{*}{\textbf{\begin{tabular}[c]{@{}c@{}}Solar \\ penetration (\%)\end{tabular}}} & \multirow{2}{*}{\textbf{\begin{tabular}[c]{@{}c@{}}\# of \\ houses\end{tabular}}} & \multicolumn{4}{c|}{\textbf{Winter}} & \multicolumn{4}{c|}{\textbf{Summer}} \\ \cline{4-11} 
 &  &  & \multicolumn{1}{c|}{\textbf{MAE (kW)}} & \multicolumn{1}{l|}{\textbf{MAPE (\%)}} & \multicolumn{1}{c|}{\textbf{PBB (\%)}} & \textbf{CRPS} & \multicolumn{1}{c|}{\textbf{MAE (kW)}} & \multicolumn{1}{l|}{\textbf{MAPE (\%)}} & \multicolumn{1}{c|}{\textbf{PBB (\%)}} & \textbf{CRPS} \\ 
 \hline
 \multirow{5}{*}{\rotatebox[origin=c]{90}{\textit{\textbf{\begin{tabular}[c]{@{}c@{}}kPF-\\AE-LSTM\end{tabular}}}}} & \textit{\textbf{15.42}} & \textit{\textbf{40}} & \multicolumn{1}{c|}{\textit{\textbf{0.24}}} & \multicolumn{1}{c|}{\textbf{2.44}} & \multicolumn{1}{c|}{\textit{\textbf{93.78}}} & \textit{\textbf{0.10}} & \multicolumn{1}{c|}{\textit{\textbf{0.18}}} & \multicolumn{1}{c|}{\textbf{2.30}} & \multicolumn{1}{c|}{\textit{\textbf{94.20}}} & \textit{\textbf{0.075}} \\ \cline{2-11}
 & \textit{\textbf{11.23}} & \textit{\textbf{50}} & \multicolumn{1}{c|}{\textit{\textbf{0.21}}} & \multicolumn{1}{c|}{\textbf{2.34}} & \multicolumn{1}{c|}{\textit{\textbf{94.45}}} & \textit{\textbf{0.07}} & \multicolumn{1}{c|}{\textit{\textbf{0.19}}} & \multicolumn{1}{c|}{\textbf{2.31}} & \multicolumn{1}{c|}{\textit{\textbf{94.00}}} & \textit{\textbf{0.07}} \\ \cline{2-11}
 & \textit{\textbf{21.23}} & \textit{\textbf{100}} & \multicolumn{1}{c|}{\textit{\textbf{0.26}}} & \multicolumn{1}{c|}{\textbf{2.51}} & \multicolumn{1}{c|}{\textit{\textbf{92.80}}} & \textit{\textbf{0.11}} & \multicolumn{1}{c|}{\textit{\textbf{0.20}}} & \multicolumn{1}{c|}{\textbf{2.40}} & \multicolumn{1}{c|}{\textit{\textbf{92.11}}} & \textit{\textbf{0.09}} \\ \cline{2-11}
 & \textit{\textbf{20.19}} & \textit{\textbf{150}} & \multicolumn{1}{c|}{\textit{\textbf{0.26}}} & \multicolumn{1}{c|}{\textbf{2.21}} & \multicolumn{1}{c|}{\textit{\textbf{92.21}}} & \textit{\textbf{0.12}} & \multicolumn{1}{c|}{\textit{\textbf{0.19}}} & \multicolumn{1}{c|}{\textbf{2.43}} & \multicolumn{1}{c|}{\textit{\textbf{92.31}}} & \textit{\textbf{0.10}} \\ \cline{2-11}
 & \textit{\textbf{21.13}} & \textit{\textbf{200}} & \multicolumn{1}{c|}{\textit{\textbf{0.27}}} & \multicolumn{1}{c|}{\textbf{2.89}} & \multicolumn{1}{c|}{\textit{\textbf{90.81}}} & \textit{\textbf{0.15}} & \multicolumn{1}{c|}{\textit{\textbf{0.23}}} & \multicolumn{1}{c|}{\textbf{2.92}} & \multicolumn{1}{c|}{\textit{\textbf{90.12}}} & \textit{\textbf{0.14}} \\ \cline{2-11}
 & \textit{\textbf{23.41}} & \textit{\textbf{210}} & \multicolumn{1}{c|}{\textit{\textbf{0.31}}} & \multicolumn{1}{c|}{\textbf{2.98}} & \multicolumn{1}{c|}{\textit{\textbf{90.17}}} & \textit{\textbf{0.15}} & \multicolumn{1}{c|}{\textit{\textbf{0.26}}} & \multicolumn{1}{c|}{\textbf{3.10}} & \multicolumn{1}{c|}{\textit{\textbf{90.91}}} & \textit{\textbf{0.13}} \\ \hline

\end{tabular}
\caption{Robustness of forecast performance of the proposed model at varying level of aggregation (number of solar sites). The results show an improvement in forecast performance compared to the individual sites (in Table\,\ref{table11}).}\label{table3}
\end{table*}
While the above study focuses on the performance of the model at individual sites, we also evaluate the model at varying levels of aggregation of the sites, which may represent the forecasting problem at the sub-station of a distribution feeder. In Table \ref{table3}, our proposed model is evaluated for six different levels of aggregation, denoted by the number of aggregated sites, each with a different aggregated solar penetration. The proposed model demonstrates improved performance at the aggregated levels, achieving less than 3\% MAPE, compared to greater than 4.75\% MAPE values at the individual site levels (in Table\,\ref{table11}). 

\subsubsection{Missing Measurements} 
\begin{figure}
    \centering
    \includegraphics[width=\columnwidth]{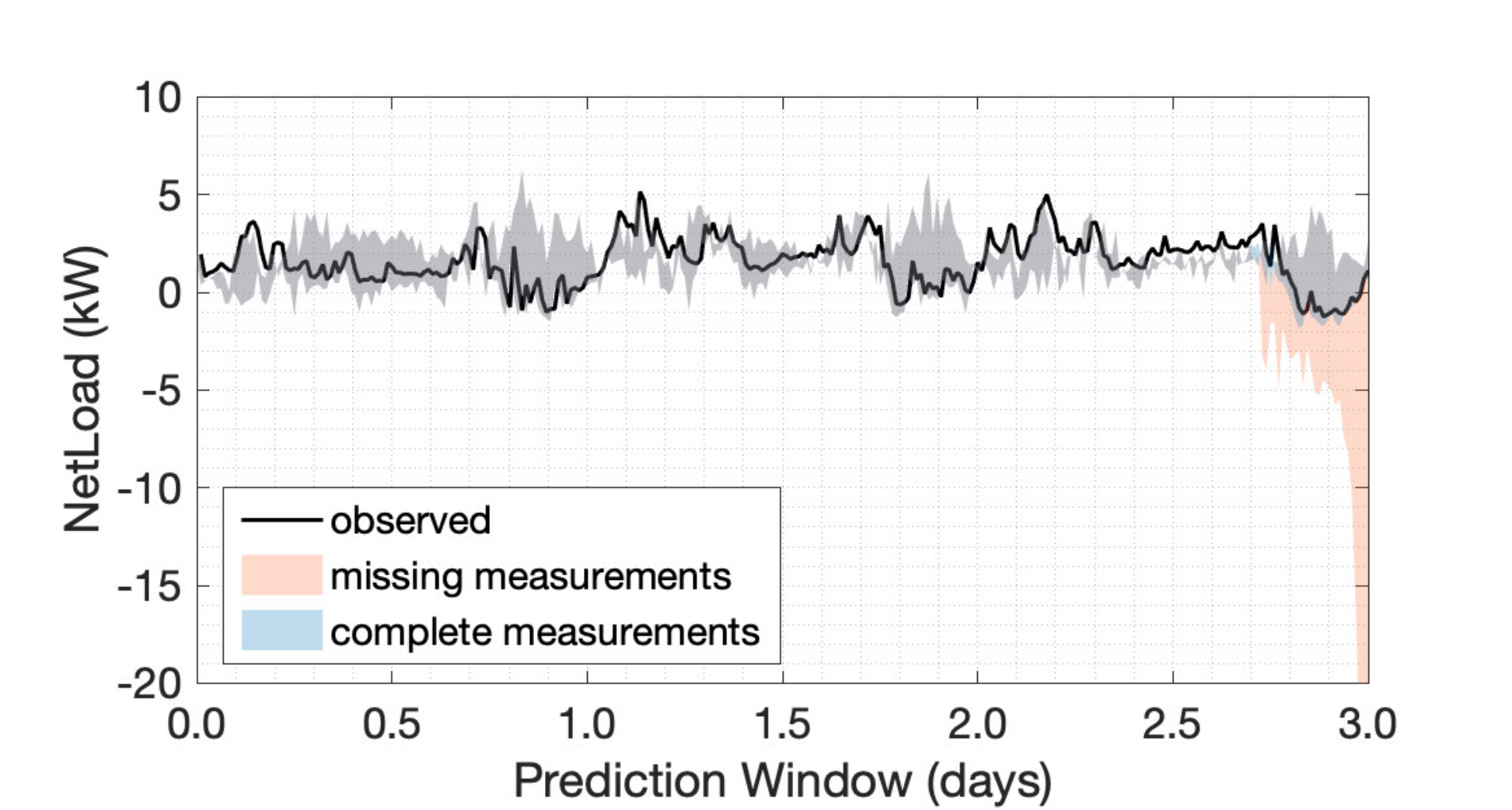}
    \caption{Impact of missing measurements (no net-load measurement beyond t=0) on the forecast performance. The `red' region denotes the 95\% confidence interval of day-ahead forecast (in rolling fashion) when the net-load measurements are missing. The `blue' region denotes the 95\% confidence interval when complete measurements are available, while the `black' line denotes the observed net-load values.}
    \label{fig3}
\end{figure}
We study the impact of missing net-load measurements on the forecasting performance. 
In Fig.\,\ref{fig3}, we studied the effect on forecasting performance, when the net-load measurements are not available for 288 consecutive time-points (or, 3 days). We assumed that the meteorological measurements continue to remain available. The `red' region shows the 95\% confidence interval of the day-ahead forecast (in a rolling fashion) when the net-load measurements are missing (i.e., unavailable during the 3 days window). The `blue' region shows the 95\% confidence interval of the day-ahead forecast when all the required measurements are updated, while the `black' line shows the observed net-load values. Even though the upper bound in `red' (missing measurements) seems to be quite consistent with the upper bound in `blue' (complete measurements), throughout the 3 days window, the lower bound in `red' seems to diverge significantly towards the end of the 3-days period, especially after 2.7 days. This shows that the proposed model, in its current architecture, may generate reliable net-load forecast for 2 days into the future, assuming meteorological data (or, forecast) are available.

\section{Conclusion}\label{sec5}
In this work, we propose a novel deep network architecture for probabilistic net-load forecasting. This proposed architecture is shown to achieve the best performance in forecast accuracy and training efficiency among the. state-of-the-art deep probabilistic forecast architectures (specifically, probabilistic convolutional-LSTM and variational RNN). We demonstrate that the proposed architecture, combined with the proposed training process, is able to generate reliable forecast performance at varying levels of behind-the-meter solar penetration (up to 50\% in annual energy), in addition to providing robust performance under different levels of aggregation, and missing measurements. 

\section*{Acknowledgment}
This work was performed jointly at Lawrence Livermore National Laboratory (under Contract DE-AC52-07NA27344) and Pacific Northwest National Laboratory (under Contract DE-AC05-76RL01830), under the auspices of the U.S. Department of Energy, supported by the Solar Energy Technologies Office (SETO) project no. 37774. The authors thank the SETO Technology Managers, Tassos Golnas and Kemal Celik, for their valuable inputs and suggestions throughout this effort.


%





\ifCLASSOPTIONcaptionsoff
  \newpage
\fi





\bibliographystyle{IEEEtran}
\bibliography{IEEEabrv,Bibliography}

\vfill



\end{document}